\documentclass{article}
\usepackage{spconf,amsmath,graphicx}
\usepackage{arydshln}
\usepackage{xcolor}
\usepackage{slashbox}
\usepackage{multirow}
\usepackage{amsfonts}
\usepackage{booktabs}
\usepackage{hyperref}

\newcommand{\ra}[1]{\renewcommand{\arraystretch}{#1}}

\title{DiariST: Streaming Speech Translation with Speaker Diarization}

\name{\begin{tabular}{c}Mu Yang$^{1\dagger}$, Naoyuki Kanda$^2$, Xiaofei Wang$^2$, Junkun Chen$^2$, Peidong Wang$^2$, Jian Xue$^2$,\\Jinyu Li$^2$, Takuya Yoshioka$^2$\end{tabular}\thanks{$^\dagger$Work performed during an internship at Microsoft.}}
\address{$^1$Center for Robust Speech Systems (CRSS), University of Texas at Dallas, Richardson, TX, USA \\$^2$Microsoft, One Microsoft Way, Redmond, WA, USA\\
{\small \tt  mu.yang@utdallas.edu, nakanda@microsoft.com}}
\begin{document}
\ninept
\maketitle
\begin{abstract}

End-to-end speech translation (ST) for conversation recordings involves several under-explored challenges such as speaker diarization (SD) without accurate word time stamps and handling of overlapping speech in a streaming fashion. In this work, we propose DiariST, the first streaming ST and SD solution. It is built upon a neural transducer-based streaming ST system and integrates token-level serialized output training and t-vector, which were originally developed for multi-talker speech recognition. Due to the absence of evaluation benchmarks in this area, we develop a new evaluation dataset, DiariST-AliMeeting, by translating the reference Chinese transcriptions of the AliMeeting corpus into English. We also propose new metrics, called speaker-agnostic BLEU and speaker-attributed BLEU, to measure the ST quality while taking SD accuracy into account. Our system achieves a strong ST and SD capability compared to offline systems based on Whisper, while performing streaming inference for overlapping speech. To facilitate the research in this new direction, we release the evaluation data, the offline baseline systems, and the evaluation code.
\end{abstract}
\section{Introduction}
\label{sec:intro}
\vspace{-.5em}

Speech translation (ST) is a task to convert speech signals into texts in other languages.
The field of ST has been extensively studied with the aim of reducing language barriers.
Traditionally, ST has been implemented by cascading two separate processes: automatic speech recognition (ASR)
and machine translation (MT) \cite{ney1999speech,matusov2005integration,post2013improved}. 
However, such a cascaded system has several limitations.

Firstly, errors originating from the ASR may propagate to the MT.
Secondly, the non-linguistic information, such as short pauses and prosody, 
may not be fully leveraged because of the text-based intermediate representation.
Lastly, the latency of the cascaded system tends to be large because 
the MT system needs to wait for the ASR result to be generated. 
To overcome these shortcomings, 
end-to-end (E2E) ST systems that directly convert speech signals into text without a separate ASR stage have been extensively studied (e.g., \cite{berard2016listen,vila2018end,sperber2020speech,radford2022robust}).
With the recent advancements in deep learning,
neural E2E ST systems
have achieved significantly better accuracy on multilingual translation using a single model \cite{barrault2023seamlessm4t},
compared to
strong cascaded systems.
Furthermore, E2E ST systems based on neural transducer~\cite{xue22d_interspeech,xue2022weakly,wang23oa_interspeech} 
have succeeded in achieving low-latency ST while maintaining higher accuracy compared to traditional, non-streaming cascaded systems.

In conversational translation scenarios,
accurately identifying the speaker of each utterance is of great importance to end users. 
This is known as the speaker diarization (SD) problem \cite{park2021review}, which has been long studied
in the context of speaker-attributed ASR \cite{fiscus2007rich,watanabe2020chime}.
For a cascaded ST system that uses separate ASR and MT components, 
conventional SD techniques employing word-level
timestamps (e.g. \cite{huang2007ibm,silovsky2012incorporation})
can be leveraged.
However, unlike ASR, it is not straightforward to obtain word-level time stamps
for E2E ST systems where the neural networks implicitly learn the mapping between input audio signals and output translated texts.
Note that, although a few techniques have recently been proposed for E2E ASR
to estimate word-level time stamps \cite{radford2022robust,bain23_interspeech},
they assume a monotonic alignment between the input and output,
which does not hold in the ST task.
Adding to the complexity of the SD,
the presence of the overlapping speech in conversations \cite{shriberg2001observations,watanabe2020chime,yu2022m2met} 
further makes 
the development of the ST systems challenging.
While there have been a lot of studies for ASR and SD,
to the best of our knowledge, no prior works have investigated the integration of ST and SD for conversational recordings.

In this work, 
we propose DiariST, the first streaming ST and SD system specifically designed for conversational recordings.
Due to the lack of prior work in this area, we first develop a new evaluation dataset for the task of translating Chinese audio to English text. This dataset, named DiariST-AliMeeting, is developed by
translating 
the Mandarin Chinese meeting recordings, AliMeeting corpus~\cite{yu2022m2met,yu2022summary},
into English.
We also establish an evaluation scheme consisting of two novel metrics, 
named 
speaker-agnostic BLEU (SAgBLEU) and speaker-attributed BLEU (SAtBLEU),
to measure the translation quality by taking the SD accuracy into account. 
We develop a streaming multi-talker ST and SD system, dubbed DiariST,
by
integrating 
the 
token-level serialized output training (t-SOT) \cite{kanda22arxiv}
and t-vector \cite{kanda22b_interspeech}, both originally developed for speaker-attributed ASR,
into the neural transducer-based ST system.
The proposed system shows a strong ST and SD capability 
compared to the offline baseline systems based on Whisper \cite{radford2022robust} 
while allowing for streaming inference, even for overlapping speech.
To facilitate the research in this new direction, 
we release the evaluation data, the offline baseline systems, 
and the evaluation code.\footnote{\url{https://github.com/Mu-Y/DiariST}}

\section{Evaluation data and metric}
\vspace{-.5em}

\subsection{DiariST-AliMeeting}
\label{sec:data}
\vspace{-.5em}

We have developed a new evaluation data, named DiariST-AliMeeting,
for the task of Mandarin Chinese audio to English text ST. 
DiariST-AliMeeting is based on the AliMeeting corpus \cite{yu2022m2met,yu2022summary}, 
which consists of Mandarin Chinese meeting recordings 
recorded by an 8-channel microphone array, as well as 
independent headset microphones (IHM) worn by each participant.
The training, development
\footnote{We renamed the ``evaluation set" in the original paper~\cite{yu2022m2met,yu2022summary} to ``development set"
to follow a standard naming convention.},
and
test sets include 
209 sessions (104 hr), 8 sessions (4 hr), and 20 sessions (10 hr), respectively.\footnote{Originally, \cite{yu2022m2met} reported 212 sessions in the training set. However, we found only 209 distributed sessions in \url{https://www.openslr.org/119/}. We report the statistics based on the distributed data.}
Each meeting session consists of a 15 to 30-min discussion by 2 to 4 participants,
and all sessions are transcribed.
Notably, the recordings contain a substantial amount of speech overlaps
with an overlapping ratio over 30\%, which makes the data challenging for ASR and SD.

In order to leverage the AliMeeting corpus for ST and SD evaluation, 
we created the English translation of 
the original Mandarin Chinese  
transcription.
We asked human translators to create 
the translation for development and test set
while we utilized GPT-4 \cite{openai2023gpt}\footnote{The API of ``gpt-4-0613'' was used.} to translate the training set.

We leveraged the audio recorded by different recording devices to create three levels of difficulties for development/evaluation sets:
(1) single distant microphone ({\bf SDM}) audio, for which
we used the first channel of the distant microphone array;
(2) mixture of IHM audio ({\bf IHM-MIX}),
where we simply mixed all the IHM signals for each meeting session;
(3) concatenation of utterances from IHM audio ({\bf IHM-CAT}), 
where the utterances are sorted based on their start times.
Among the three levels, SDM audio presents the greatest challenge due to its inclusion of overlapping speech and relatively low Signal-to-Noise Ratio (SNR).
IHM-MIX is less challenging compared to SDM, as it features a higher SNR while still containing overlapping speech.
IHM-CAT is the least challenging because of the high-SNR signals without overlapping speech.
Note that, even though IHM-CAT is the least challenging of the three, performing SD on IHM-CAT audio is non-trivial due to the frequent speaker turns.

In addition, we divided the original long-form recordings into shorter segments, ranging from 3 to 6 minutes, for both the development and evaluation sets. 
This is because the BLEU score \cite{papineni2002bleu} is computed based on the precision of N-grams,
which tends to be over-estimated
when the length of the ground truth translation
is long.
As a result, we created the development and evaluation sets
containing 76 mini-sessions and 195 mini-sessions,
respectively.

\vspace{-.5em}
\subsection{SAgBLEU and SAtBLEU}
\vspace{-.5em}
The ST quality of conventional single-talker speech-to-text systems is usually evaluated by the BLEU score \cite{papineni2002bleu}. In the multi-talker case, it is desired to evaluate the ST quality, considering the SD errors as well. Such an evaluation scheme has been studied in the context of ASR, where prior works proposed several extensions to word error rate (WER) such as speaker-attributed WER \cite{fiscus2007rich} and concatenated minimum permutation WER (cpWER) \cite{watanabe2020chime}. 
In this work,
we propose speaker-agnostic BLEU (SAgBLEU) and speaker-attributed BLEU (SAtBLEU) to assess the ST performance without 
and with considering the SD quality.
Our proposed metric is inspired by the cpWER,
which does not require precise time stamps for each word.

The evaluation problem is formulated as follows: in each mini-session,
we assume that reference utterances 
are indexed in the order of their start times,
and the reference translation for $i$-th utterance is denoted as $r_i^{s_i}$,
where $s_i\in\mathcal{S}$ represents the speaker index of $i$-th utterance,
and $\mathcal{S}$ denotes the set of unique reference speaker indices in the mini-session.
We also assume that the ST and SD system outputs 
a hypothesis containing multiple translated utterances,
where the translated utterances are indexed
in the order of their start times.
We denote the $j$-th hypothesis utterance as $h_j^{p_j}$,
where
$p_j\in\mathcal{P}$ denotes 
the predicted speaker index for the $j$-th utterance,
and $\mathcal{P}$ represents the set
of unique predicted speaker indices in the mini-session.

\textbf{SAgBLEU} computes the BLEU score without considering the SD errors.
Here, 
we first form a concatenated reference translation  $r=r_{1}^{s_1}r_{2}^{s_2}...r_{M}^{s_M}$,
where $M$ denotes the total number of reference utterances in this mini-session.
We also form a concatenated hypothesis translation
$h=h_{1}^{p_1}h_{2}^{p_2}...h_{N}^{p_N}$, where $N$ denotes
the total number of hypothesis utterances.
The BLEU score is then computed at the corpus level by aggregating $h$ and their corresponding $r$ across all mini-sessions.\footnote{SacreBLEU~\cite{post2018call} version 2.3.1 is used in our implementation.}

\textbf{SAtBLEU} computes the BLEU score while considering the SD errors.
Here, we
first form speaker-wise references.
More specifically,
a speaker-wise reference $r^s$ for speaker $s\in\mathcal{S}$ is 
made by concatenating $r_i^{s_i}$ for all $i\in M$ where $s_i=s$.
In the same way, 
a speaker-wise hypothesis 
$h^p$ for speaker $p\in\mathcal{P}$
can be made by concatenating $h_j^{p_j}$ for all $j\in N$ where $p_j=p$.
Then, we find the speaker permutation 
that maximizes the BLEU score between
speaker-wise hypotheses
 $\{h^p\}_{p \in \mathcal{P}}$
and the
 speaker-wise references $\{r^s\}_{s\in \mathcal{S}}$.
Note that, when the numbers of speakers in $\mathcal{S}$ and $\mathcal{P}$ do not match, empty (NULL) strings are added to the speaker-wise hypotheses or references to ensure the matching number of speakers. 
Finally, we compute corpus-level BLEU score \cite{papineni2002bleu,post2018call} by aggregating the speaker-wise hypotheses and the corresponding speaker-wise references from all mini-sessions, with the identified session-level permutations applied.

\section{DiariST: Streaming ST and SD system}
\label{sec:proposed}
\vspace{-.5em}

\subsection{ST with neural transducer}
\vspace{-.5em}

In this work, we adopt neural transducer as the backbone
of our ST model \cite{xue22d_interspeech}.
Unlike many E2E ST methods based on attention encoder-decoder (e.g. \cite{radford2022robust,barrault2023seamlessm4t}),
a neural transducer-based ST model has the advantage of streaming inference
while naturally performing word reordering.
It was shown that the neural transducer-based ST model outperforms offline cascaded ST systems while maintaining the streaming inference capability. For additional details, please refer to \cite{xue22d_interspeech}.

\vspace{-.5em}
\subsection{Streaming multi-talker ST with t-SOT}
\vspace{-.5em}

The idea of t-SOT was originally proposed in the streaming ASR task  for handling overlapping speech \cite{kanda22arxiv}, which is common in conversational recordings
\cite{shriberg2001observations,watanabe2020chime,yu2022m2met}.
In the t-SOT framework, given audio with multiple active speakers, the ASR model is trained to generate
the transcriptions of all speakers in chronological order based
on the end time of each token.
To distinguish the transcriptions of different speakers in the regions of overlapping speech,
a special separator symbol $\langle \mathrm{cc}\rangle$ is inserted 
when the adjacent tokens belong to different speakers.
In inference, the ASR model generates
a sequence of tokens including
$\langle \mathrm{cc}\rangle$, which is subsequently deserialized into
multiple sequences that are supposed to contain 
non-overlapping tokens.
The t-SOT framework achieved highly accurate multi-talker speech recognition performance
while retaining the streaming inference capability \cite{kanda22arxiv}.

In this work, we adopt the t-SOT framework to handle overlapping speech
in the ST task
for its simplicity and proven accuracy on the ASR task.
The challenge of using the t-SOT framework for ST lies in the fact that,
unlike ASR,
the physical emission time for each word is unknown, making it difficult to sort the tokens in chronological order.

To overcome this challenge, we propose a two-step approach. First,
we train a neural transducer-based single-talker ST model.
Next, this single-talker ST model is used to generate the Viterbi alignment
between the training audio sample and ground-truth translation.
In this way, we are able to obtain the model-based token emission time
for the training data,
which is then used to sort the tokens for the training of the t-SOT-based ST model.

\vspace{-.5em}
\subsection{Streaming SD with t-vector}
\vspace{-.5em}

The challenge of SD in the context of ST is that
it is difficult to obtain word-level time stamps,
which are often the basis
for conventional SD with ASR systems \cite{huang2007ibm,silovsky2012incorporation}.
SD becomes further challenging when handling overlapping speech,
which contaminates one speaker's representation with other speakers' characteristics.

To tackle these challenges,
we leverage t-vector \cite{kanda22b_interspeech}, a token-level speaker embedding, which was
originally developed for neural transducer-based ASR.
The t-vector model comprises a speaker encoder and a speaker decoder. 
The speaker encoder
consists of a Res2Net module \cite{zhou2021resnext,liu2022microsoft} followed by a stack of
 multi-head attention (MHA) layers. 
The MHA layers in the speaker encoder work in tandem 
with the ASR encoder on a frame-by-frame basis,
where the key and query of the MHA in the ASR encoder 
is reused as the key and query of the MHA in the speaker encoder.
On the other hand, the speaker decoder consists of two long short-term memory (LSTM) layers 
and generates a t-vector for each token
given the speaker encoder output and the target token embedding.
During inference, t-vectors are estimated in a streaming fashion, in parallel with
token estimation. 
These t-vectors are then fed into the online clustering algorithm
for speaker diarization.
Refer to \cite{kanda22b_interspeech} for more details.

\section{Experiments}
\vspace{-.5em}

\subsection{Configurations of offline baseline systems}
\vspace{-.5em}

We first benchmark the performance of representative ST and SD methods on our proposed evaluation dataset, DiariST-AliMeeting (Section \ref{sec:data}).
These serve as the reference points to indicate the dataset's level of difficulty.
For this purpose,
we developed two offline ST and SD systems
using Whisper as the ST backbone~\cite{radford2022robust}.
Specifically, all our experiments employed the ``Whisper Small'' model (244M parameters), whose parameter size is the closest to our neural transducer-based ST model (216M parameters).

The first offline system executed the clustering-based
SD, followed by a Whisper-based ST for each region detected by SD.
We call it ``{\bf SD$\rightarrow$ST system}''.
More specifically,
we first 
applied voice activity detection (VAD) \cite{speechbrain}.
Next, we uniformly segmented the speech region by a 1.2-second sliding window with a 0.6-second shift, from which we
extracted ECAPA-TDNN-based speaker embedding~\cite{DBLP:conf/interspeech/DesplanquesTD20,speechbrain}.
We then applied
normalized maximum eigengap-based spectral clustering (NME-SC)~\cite{park2019auto}
with the maximum number of speakers set to 6.
Finally, we segmented the audio based on the SD result and applied
Whisper ST for each segment independently.

The second offline baseline system first applied
the Whisper-based ST on the long-form audio. 
Then the clustering-based SD was applied based on the segmentation information estimated by Whisper.
We call it ``{\bf ST$\rightarrow$SD system}''.
Specifically, we applied Whisper-based ST for the long-form audio
with a sliding window of 30 seconds \cite{radford2022robust}. 
We turned off the context biasing from the hypothesis of the prior recognition window
because we found it sometimes produced repetitive hallucination errors.
Whisper ST produces a list of \{translation of utterance, start time, end time\}.
We then extracted ECAPA-TDNN-based speaker embedding based on the start time and end time of each translation, and applied NME-SC-based clustering
with a maximum number of speakers of 6.

\vspace{-.5em}
\subsection{Configurations of streaming ST and SD systems}
\vspace{-.5em}

We developed the streaming ST and SD system based on the neural transducer with
t-SOT and t-vector, as proposed in Section \ref{sec:proposed}.
For the ST model, we used Conformer \cite{gulati2020conformer} based transducer
 with chunk-wise look-ahead \cite{chen2021developing}.
 We set the chunk size to be 1,000 msec.
The encoder of the ST model consists of two convolutional layers followed by
18 Conformer layers,
where
each Conformer layer consisted
of a 512-dim MHA with 8 heads
and two 3,072-dim point-wise feedforward layers. 
The prediction
network consisted of 2 layers of 1,024-dim LSTM. 
5,854 word
pieces plus blank, $\langle \mathrm{eos}\rangle$\cite{wang23oa_interspeech} and $\langle \mathrm{cc}\rangle$ tokens
were used as the recognition units. 
The input feature was an 80-dim log mel-filterbank
extracted for every 10 msec. 
Furthermore, the encoder of the t-vector extractor consisted of
Res2Net \cite{zhou2021resnext,liu2022microsoft} followed by 18 layers of 128-dim 8-head MHA.
The decoder of the t-vector extractor 
consisted of a 2-layer 512-dim LSTM. 

\noindent \textbf{Training of the t-SOT-based streaming ST model.} The streaming ST model was pre-trained on our
in-house data which contains \{audio, English-text\} pairs originating
from 14 languages (ar, de, es, et, fr, it, ja, nl, pt, ru, sl, sv, zh, en).
Subsequently, we fine-tuned the ST model on the training data from DiariST-AliMeeting.
This fine-tuning was informed by our preliminary experiment 
where we observed that the t-SOT framework became effective for real-world evaluation data 
only after being fine-tuned with a modest amount of real-world training data,
as observed in multi-talker ASR experiments \cite{kanda2021large,yang2023simulating}. \footnote{In multi-talker ASR experiments, such a fine-tuning significantly improved the overlapping speech handling even for unseen domain \cite{yang2023simulating,li23o_interspeech}. }
The pre-training data contains approximately 1,000 hours of audio for each language, leading to 14,000 hours of data in total.
We use multilingual pre-training data 
because 
(i) a multilingual model showed better accuracy than a monolingual model \cite{papi2023token}
and (ii) our preliminary experiments showed its efficacy in training the t-vector model using the VoxCeleb dataset, which also contains multi-lingual recordings.
To train the t-SOT model, training instances are constructed by
randomly sampling 1 to 5 samples from the training data,
and mixing them by randomly delaying each sample
under the constraint of having up to 2 active speakers at the same time frame.
We performed
600K training iterations with 16 GPUs, 
where each mini-batch contains multiple samples such that
the \#\_of\_frames $\times$ \#\_of\_labels roughly equals to 600,000. 
An AdamW optimizer was used with a linear decay learning rate schedule, peaking at 
3e-4 after 75,000 warm-up iterations.
After pre-training,
we fine-tune the ST model for
2,500 iterations with 8 GPUs on the training data from DiariST-AliMeeting.
An AdamW optimizer
 with a linear decay learning rate schedule with a peak learning
 rate of 3e-5 was used.

\noindent \textbf{Training of t-vector}. After training the streaming ST model, we trained the t-vector model using VoxCeleb \cite{nagrani2017voxceleb,chung2018voxceleb2}, following the same training configuration
used in \cite{kanda22b_interspeech}. 
Training instances are constructed in the same way as those used for t-SOT-based streaming ST model training.
We performed
125K training iterations with 16 GPUs, 
each of which consumed mini-batches of 6,000 frames.
We used an AdamW optimizer
with a linear decay learning rate schedule with a peak learning
rate of 2e-4 after 12,500 warm-up iterations.

Finally, to highlight the importance of t-SOT based multi-talker modeling, 
we trained a single-talker streaming ST model with t-vector 
as a baseline system. 
This model was trained by using a configuration nearly identical to that of
the t-SOT-based model except that the pre-training of ST was conducted 
without multi-talker data simulation,
and the fine-tuning of ST was conducted on the original IHM and SDM utterances without mixing. 
T-vector model was then trained on top of this single-talker ST model with the same training configuration.

\begin{table}[t]
\ra{0.9}
\tabcolsep = 1.2mm
  \caption{SAgBLEU, DER and SAtBLEU of the {\bf offline} ST and SD systems based on Whisper.}
  \label{tab:offline}
 \vspace{2mm}
  \centering
{   \footnotesize
\begin{tabular}{@{}cccccccc@{}}
\multicolumn{8}{c}{(a) SAgBLEU {\scriptsize ($\uparrow$)}} \\
\toprule
& \multicolumn{3}{c}{dev} && \multicolumn{3}{c}{test} \\ \cmidrule{2-4} \cmidrule{6-8}
System  & IHM-CAT & IHM-MIX & SDM  && IHM-CAT & IHM-MIX & SDM \\ 
\midrule
SD$\rightarrow$ST & 13.56   & 10.71   & 8.67 && 14.47   & 10.40   & 7.88 \\
ST$\rightarrow$SD & {\bf 15.42}   & {\bf 14.48}   & {\bf 9.81} && {\bf 18.45}   & {\bf 13.68}   & {\bf 10.51} \\ 
\bottomrule

\\
\multicolumn{8}{c}{(b) DER {\scriptsize ($\downarrow$)}} \\
\toprule
& \multicolumn{3}{c}{dev} && \multicolumn{3}{c}{test} \\ \cmidrule{2-4} \cmidrule{6-8}
System  & IHM-CAT & IHM-MIX & SDM  && IHM-CAT & IHM-MIX & SDM \\ 
\midrule
SD$\rightarrow$ST & 8.23   & 21.75   & 28.77  && 9.22   & 22.39    & 26.71 \\
ST$\rightarrow$SD & {\bf 3.70}   & {\bf 19.01}   & {\bf 23.57}  && {\bf 3.15}   & {\bf 18.50}   & {\bf 24.49} \\
\bottomrule

\\
\multicolumn{8}{c}{(c) SAtBLEU {\scriptsize ($\uparrow$)}} \\
\toprule
& \multicolumn{3}{c}{dev} && \multicolumn{3}{c}{test} \\ \cmidrule{2-4} \cmidrule{6-8}
System  & IHM-CAT & IHM-MIX & SDM  && IHM-CAT & IHM-MIX & SDM \\ 
\midrule
SD$\rightarrow$ST & 11.58   & 9.21   & 7.39 && 12.94   & 9.31    & 6.80 \\
ST$\rightarrow$SD & {\bf 13.26}   & {\bf 11.96}   & {\bf 8.21} && {\bf 16.81}   & {\bf 12.10}   & {\bf 9.02} \\
\bottomrule
  \end{tabular}
  }
\end{table}

\vspace{-.5em}
\subsection{Results}
\vspace{-.5em}
\subsubsection{Comparison of the offline ST and SD systems}
\vspace{-.5em}

Table \ref{tab:offline} shows the SAgBLEU and SAtBLEU results for the two Whisper-based offline systems.
Because these two systems provide utterance-level timestamps, we also computed
diarization error rate (DER) with a tolerance collar of 0.25 seconds.

From Table \ref{tab:offline} (a) and (b),
we first observed that the ST$\rightarrow$SD system outperformed
the SD$\rightarrow$ST system across all conditions in SAgBLEU and DER.
The better result on SAgBLEU can be attributed to the fact that
the ST$\rightarrow$SD system can leverage long-context information up to 30 second
in the ST task.
On the other hand, the SD$\rightarrow$ST system does not have this advantage as the input audio is segmented before ST processing.
For DER, the substantially better results by  
the ST$\rightarrow$SD system may be due to its utilization of the linguistic information 
for segmentation, consistent with the findings
in the ASR and SD works~\cite{park2020speaker,xia2021turn,kanda2022transcribe}.

As expected from the results of SAgBLEU and DER, 
the ST$\rightarrow$SD system showed a significantly better SAtBLEU score
compared to 
the SD$\rightarrow$ST system for all conditions as shown in Table \ref{tab:offline} (c).
However, we noticed that 
the correlation between DER and \{SAgBLEU - SAtBLEU\} is not very high
although both are indicators of SD accuracy.
For example, the difference between SAgBLEU and SAtBLEU on IHM-CAT dev set for 
the ST$\rightarrow$SD system is 2.16 (= 15.42 - 13.26) points while
that on IHM-MIX dev set is 2.52 (= 14.48 - 11.96) points.
This difference does not correlate well with the DER difference of
3.70\% vs. 19.01\%.
One possible reason may be that DER is impacted by many missing voice activities for overlapping speech 
in IHM-MIX and SDM sets
while \{SAgBLEU - SAtBLEU\} is not impacted by missing translations.

Finally, we observed a substantial degradation in the SAtBLEU scores from IHM-CAT to IHM-MIX,
and from IHM-MIX to SDM. 
The former indicates the impact of overlapping speech, and the latter indicates the impact
of the low SNR signals.

\begin{table}[t]
\ra{0.9}
\tabcolsep = 0.7mm
  \caption{SAgBLEU and SAtBLEU of the {\bf streaming} ST and SD systems. The system with t-SOT ST and t-vector is our proposed system, DiariST. Note that DER was not computed because the system does not produce time stamps for each word or utterance.}
  \label{tab:streaming}
 \vspace{2mm}
  \centering
{   \footnotesize
\resizebox{1.0\columnwidth}{!}{
\begin{tabular}{@{}cccccccc@{}}
\multicolumn{8}{c}{(a) SAgBLEU {\scriptsize ($\uparrow$)}} \\
\toprule
& \multicolumn{3}{c}{dev} && \multicolumn{3}{c}{test} \\ \cmidrule{2-4} \cmidrule{6-8}
System  & IHM-CAT & IHM-MIX & SDM  && IHM-CAT & IHM-MIX & SDM \\ 
\midrule
1spk ST  &  15.45 & 14.73 & 12.21                   && {\bf 20.63} & 16.15 & 13.46  \\
t-SOT ST &  {\bf 18.45} & {\bf 16.05} & {\bf 13.47} && 19.80 & {\bf 16.62} & {\bf 13.52} \\
\bottomrule
\\
\multicolumn{8}{c}{(b) SAtBLEU {\scriptsize ($\uparrow$)}} \\
\toprule
& \multicolumn{3}{c}{dev} && \multicolumn{3}{c}{test} \\ \cmidrule{2-4} \cmidrule{6-8}
System  & IHM-CAT & IHM-MIX & SDM  && IHM-CAT & IHM-MIX & SDM \\ 
\midrule
1spk ST + d-vec &  12.54 & 12.01 & 9.67 && {\bf 17.38}& 13.39& 10.77 \\
1spk ST + t-vec &  12.43 & 12.00 & 10.01 && 16.99& 13.22& 10.91 \\  \hdashline[1pt/2pt]\hdashline[0pt/1pt] 
t-SOT ST + d-vec &  14.61 & 13.15 & 10.78 && 16.67 & 13.85 & 10.76 \\
t-SOT ST + t-vec & {\bf 15.16} & {\bf 13.91} & {\bf 11.44} && 17.07 & {\bf 14.76} & {\bf 11.67} \\
\bottomrule
  \end{tabular}
  }
  }
\end{table}

\vspace{-.5em}
\subsubsection{Comparison of the streaming ST and SD systems}
\vspace{-.5em}

Table \ref{tab:streaming} shows the comparison of various streaming ST and SD systems.
For the comparison, we experimented with single-speaker (1spk) ST and t-SOT ST, in combination with either d-vector or t-vector. 
Here, d-vector was extracted by a 0.8-second window based on the token emission time from the ST models.
We used Res2Net trained by VoxCeleb \cite{liu2022microsoft} as the d-vector extractor, 
which was also used for the training
target of t-vector.

First, from Table \ref{tab:streaming} (a) we can observe that t-SOT ST surpassed  1spk ST in terms of SAgBLEU
for 5 out of the 6 tested conditions with IHM-CAT test set as the exception.
It is noteworthy that t-SOT ST achieved significantly better SAgBLEU for IHM-MIX and SDM for both development and test sets, which
are the real conversational recordings with speech overlaps.
We also observed that the degradation of SAgBLEU from IHM-CAT to IHM-MIX conditions was
significantly smaller compared to the offline systems that do not handle overlapping speech.
For example, t-SOT ST showed a 16.1\% (19.80 to 16.62) degradation from IHM-CAT to IHM-MIX on the test set
while 
the ST$\rightarrow$SD system showed a 25.9\% (18.45 to 13.68) degradation in the same condition.
This indicates that t-SOT ST effectively handled overlapping speech based on the $\langle \mathrm{cc}\rangle$ token.

From Table \ref{tab:streaming} (b), we observed that
the combination of t-SOT ST and t-vector achieved the best result for 5 out of the 6 conditions. 
Particularly, it achieved the best results on IHM-MIX and SDM.
We observed that t-vector becomes specifically effective when combined with t-SOT ST.
For 1spk ST, t-vector was competitive with d-vector.
We believe that it is because t-vector can better utilize $\langle \mathrm{cc}\rangle$ token emitted
by the t-SOT ST model as the clue to estimate speaker embeddings.
Overall, our proposed system, DiariST, achieved a strong SAtBLEU score
compared to the offline baselines as well as the naive streaming ST system based on 1-spk ST and d-vector.

\vspace{-.5em}
\section{Conclusion}
\vspace{-.5em}

This paper presented DiariST, 
the first streaming ST and SD system 
built upon t-SOT and t-vector.
We also presented a new dataset DiariST-AliMeeting,
and new evaluation metrics of SAgBLEU and SAtBLEU
to comprehensively assess the quality of ST in conjunction with the SD task.
From the experiments on DiariST-AliMeeting, we showed that the proposed system 
can achieve a strong ST and SD
capability compared to various baseline systems
while maintaining the streaming inference for overlapping speech.

\newpage

\bibliographystyle{IEEEtran}
\bibliography{mybib}

\end{document}